\documentclass[conference]{IEEEtran}
\IEEEoverridecommandlockouts
\usepackage{cite}
\usepackage{amsmath,amssymb,amsfonts}
\usepackage{algorithmic}
\usepackage{graphicx}
\usepackage{textcomp}
\usepackage{amsmath}
\usepackage{verbatim}
\usepackage{xcolor}
\usepackage{multirow}
\usepackage{subfig}
\usepackage{hyperref} 
\usepackage{fancyhdr}
\usepackage{newunicodechar}
\newunicodechar{ﬁ}{fi}
\newunicodechar{ﬀ}{ff}
\usepackage[utf8x]{inputenc}

\def\BibTeX{{\rm B\kern-.05em{\sc i\kern-.025em b}\kern-.08em
    T\kern-.1667em\lower.7ex\hbox{E}\kern-.125emX}}
\begin{document}

\title{Online Self-Evolving Anomaly Detection in Cloud Computing Environments 
}

\author{
  \IEEEauthorblockN{
    \textit{Haili Wang}\IEEEauthorrefmark{1},
    \textit{Jingda Guo}\IEEEauthorrefmark{1},
    \textit{Xu Ma}\IEEEauthorrefmark{1},
    \textit{Song Fu}\IEEEauthorrefmark{2},
    \textit{Qing Yang}\IEEEauthorrefmark{2} and 
    \textit{Yunzhong Xu}\IEEEauthorrefmark{1}
  }
  \IEEEauthorblockA{
    Department of Computer Science and Engineering \\
    University of North Texas\\
    Denton, Texas, USA
  }\IEEEauthorblockA{
    \IEEEauthorrefmark{1}\{HailiWang, JingdaGuo, XuMa, YunzhongXu\}@my.unt.edu,
    \IEEEauthorrefmark{2}\{Song.Fu,\ Qing.Yang\}@unt.edu,
  }
}

\maketitle
\thispagestyle{fancy}
\pagestyle{fancy}

\begin{abstract}
Modern cloud computing systems contain hundreds to
thousands of computing and storage servers. Such a scale, combined with ever-growing system complexity, is causing a key challenge to failure and resource management for dependable cloud computing.
Autonomic failure detection is a crucial
technique for understanding emergent, cloud-wide phenomena and self-managing cloud resources for system-level dependability assurance. To detect failures, we need to monitor the cloud
execution and collect runtime performance data. These data are usually unlabeled, and thus a prior failure history is not always available in production clouds.
In this paper, we present a \emph{self-evolving 
anomaly detection} (SEAD) framework for cloud dependability assurance.
Our framework self-evolves by recursively exploring newly verified anomaly records and continuously updating the anomaly detector online. As an distinct advantage of our framework, cloud system administrators only need to check a small number of detected anomalies and their decisions are leveraged to update the detector. Thus, the detector evolves following the upgrade of system hardware, update of software stack, and change of user workloads. Moreover, we design two types of detectors, one for general anomaly detection and the other for type-specific anomaly detection. With the help of self-evolving technique, our detectors can achieve 88.94\% in sensitivity and 94.60\% in specificity on average, which makes them suitable for real-world deployment. 

\end{abstract}

\begin{IEEEkeywords}
Cloud computing; Dependable systems; Online learning, Anomaly detection; Autonomic management
\end{IEEEkeywords}

\section{Introduction}

Cloud computing is widely used in almost all aspects of our daily life~\cite{zhang2010cloud}, from social media, online shopping, on-demand movies, photo storage, document editing to scientific computing, big data processing, and smart cities (e.g., autonomous vehicles,  smart transportation and more). Production cloud systems, such as Amazon Web Services, Google Cloud Platform and Microsoft Azure, are both economically successful and technically popular.    


   Despite great efforts on the design of ultra-reliable
   components, the increase of cloud size and complexity has outpaced the improvement of component reliability. Results from recent studies show that
   the reliability of existing data centers and cloud computing systems is constrained by a mean time between failure (MTBF) on the order of 10 - 100 hours. Failure occurrence as well as its impact on cloud performance and operating costs becomes an increasingly important concern to cloud system operators and cloud service providers. 
   
   Anomaly detection is an important failure management technology for computer systems. It detects anomalous system/component behaviors and possible failures by analyzing history behaviors and execution states. Anomaly detection in cloud computing systems provides a cost-effective mechanism for resource allocation, virtual machine/container scheduling, and cloud maintenance.

   During cloud operations, a large amount of monitoring data is collected to track the cloud's operational status. Software log files, system audit events, and network traffic statistics are typical examples of such measurements. These data
   provide valuable information about the cloud's
   health. A failure occurrence scatters its trace in the measurement data so that we can interpret the data to identify a system/component failure. However, the cloud
   measurements usually contain enormous numbers of attributes and continuous monitoring leads to the overwhelming data volumes. It is impossible to manually infer the cloud status from those measurements. Another
   challenge of anomaly detection from measurement data
   originates from the dynamics of cloud computing systems. It is common in those systems that user behaviors and servers' loads are always changing. The cloud hardware and software components are also frequently replaced or updated. This requires the anomaly detection mechanisms distinguish the normal cloud variation and real failures.

   The conventional methods of anomaly detection rely on statistical learning algorithms to approximate the dependency of failure occurrences on various performance attributes; see for a comprehensive review and for examples. The underlying assumption of these methods is that the training dataset is labeled, i.e. for each measurement used to train a failure detector, the designer knows if it corresponds to a normal execution state or a failure. However, the labeled data are not always available in real-world cloud computing environments.
   
   Moreover, these methods usually adopt an offline-training-and-online-detection scheme. Specifically, an anomaly detector is trained offline using a vast amount of labeled data. Then the detector is used for online anomaly detection. Once deployed, the detector is seldom changed unless the detection
   accuracy is worse than expected (i.e., below some predefined accuracy threshold), which leads to a retraining of the detector offline and replacing the online detector by the new one. As a result, the performance of anomaly detection fluctuates, that is
   the detection accuracy drops as time goes on until replacement by the retrained detector, which causes an abrupt improvement of the detection accuracy. After that, the cycle repeats. As the retraining phase is time-consuming, the badly-behaved detector remains in service for a long while, which causes more system/component failures undetected and false alarms. The fluctuating detection performance affects the efficacy of system monitoring and resource management as well. 
   
   In this paper, we address these issues by presenting a {\em self-evolving anomaly detection framework}, named SEAD. SEAD adopts a novel ... and adaptive anomaly detection approach. Specifically, SEAD does not require a prior failure history. It continuously monitors the cloud execution and collects runtime performance data. To tackle the high dimensionality of cloud performance metrics, SEAD extracts the most relevant metrics for anomaly detection. It describes the cloud performance data using .... and employs ... to detect possible failures. As the detections get verified by the cloud operators, they are confirmed as either true failures with failure types or normal states. SEAD adapts its anomaly detector by recursively learning from these newly verified detection results. The ... are updated to refine future detections. In addition, the cloud operators can report observed but undetected failure records to the SEAD detector, which exploits these records to identify new types of failures.

Considering that limitations are originating from
high complexity and dynamicity of the cloud computing system, the autonomic failure management is an effective approach in enhancing cloud dependability. Specifically, anomaly detection is one of the most important failure management techniques in cloud computing. It detects anomalous behaviors and possible failures in a cloud system by analyzing the previous system records and running states. Anomaly detection in cloud computing system provides a reliable failure management method for resource allocation, virtual machine reconfiguration and cloud maintenance.

Considering that manual analysis is impossible for large amount of cloud data, statistical learning based anomaly detection has prevailed on this field. Most related studies identify potential anomalies in cloud computing system by examining historical system records. Such an anomaly detection method requires large amount of normal system records and anomalies to successfully identify an unknown behavior. However, only limited amount of records are labeled in most cases with no prior failure history, especially for new clouds. 

\begin{figure}
    \centering
    \includegraphics[width=0.9\linewidth]{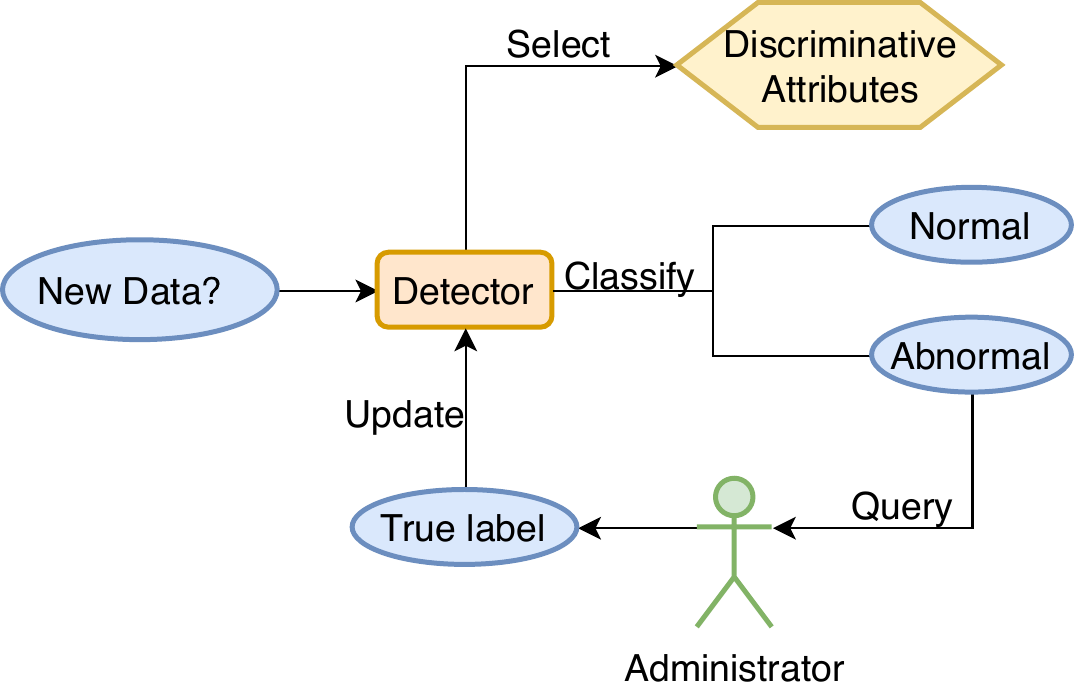}
    \label{fig:framework}
    \caption{Our self-evolving framework. Once a batch of new records is generated, we feed the new data to the detector to obtain classification. The predicted anomalies are then sent to administrator for correct labels. The verified records will be used to update our detector in next iteration. The detector can also select the most distinguishing attributes for further studies. }
\end{figure}

Differing from the existing anomaly detection approaches, we exhibit a self-evolving framework for anomaly detection, as shown in Fig \ref{fig:framework}, which requires no previous identified records, and thereby is able to learn the data distribution incrementally and improve the detection performance progressively. We build our new customized detectors on the basis of traditional statistical learning. Our final self-evolving framework can achieve 99.94\% in detection sensitivity and 94.60\% in detection specificity, indicating that our framework is practical for real cloud computing environments.

In summary, the main contributions of this paper include:
\begin{itemize}
  \item We present a self-evolving framework for cloud computing environment anomaly detection with no requirements on previously labeled records and can learn from newly generated records incrementally. Moreover, only 21.07\% of records in the system needs labeling in the whole process. Comparing with current approaches, our framework can greatly reduce the administrator's work.
  \item In pursuit of online learning, we introduce an elegant yet efficient detector. For convenience, we adapt stochastic gradient descent method to solve our model.
  \item While most of anomaly detection methods only classify a record as either normal or anomaly, which ignore the difference in anomalies, our framework can detect abnormal records with their particular anomalies. 
  \item By adding a sparsity regularization term to our detector model, it is possible for our framework to select the most distinguishing attributes for further researches.
  \item Extensive experiments show that the proposed self-evolving framework achieves better performance than the previous ones in cloud computing environment anomaly detection.
\end{itemize}

The remaining paper is organized as follows: In Section II, we briefly review related works on cloud computing anomaly detection. Section III presents the pipeline of our self-evolving framework and comprehensively introduces our novel detectors. Section IV presents experimental results and analysis. Lastly, section V concludes the paper and discusses future works.

\section{Related Work}
Theories and practices that apply machine learning methods on cloud anomaly detection have been studied for many years. An early practice of learning cloud anomaly detection is  to apply Naive Bayes \cite{rish2001empirical} for detecting cloud anomaly \cite{smith2010anomaly}. Combined with feature transformation algorithms, the work in \cite{smith2010anomaly} achieved a satisfying performance.
Following the same strategy, a distributed approach \cite{palmieri2014distributed} adopted Independent Component Analysis (ICA) \cite{hyvarinen2004independent} instead of Principal Component Analysis (PCA) \cite{jolliffe2011principal}. Besides, a decision tree model is employed for classifying anomaly. The general pipeline (composed of two modules: an unsupervised feature transformation module that \textbf{preprocesses} the collected data and a supervised classification module that detects anomaly) has also been used in numerous works for disk failure analysis, network security, cloud computing detection, and more. Nevertheless, simply detecting cloud computing environment anomaly by a supervised algorithm has obvious limitations. The increasing unlabeled data logs makes this approach impractical for real online detection applications.  

Considering the lack of prior label (\textbf{ing}) knowledge, Pannu et al. \cite{pannu2012aad}, managed to use unsupervised methods (one class support vector machine \cite{amer2013enhancing}) in cloud computing anomaly detection. This simple yet effective strategy allows the adaptive system to surpass the previously published work in real word applications, such as \cite{wang2011statistical,guan2012ensemble}. The success of \cite{pannu2012aad} chiefly comes from (i) no introduction of prior label \textbf{ing} knowledge in model learning stage \textbf{word cut?}; (ii) the high-capacity of one class SVM. Analogous to \cite{pannu2012aad} that applied unsupervised algorithms in the whole system, \cite{fu2012hybrid,huang2016related} also employed support vector data description (SVDD for short) \cite{tax2004support} to detect ICA/Kernel-PCA transformed data. Empirical experiments demonstrated that \cite{fu2012hybrid} improved the detection sensitivity by 19.6\%, over a strong baseline of Bayesian predictors and decision trees ensemble \cite{guan2012ensemble}. 

With the development of machine learning, a trend of using self-evolving algorithms to detect cloud computing anomaly is emerged \cite{pannu2012self}. The supervised detector self-evolves in a manner of recursively feedback high-confidence results for future anomaly identification. Hence, self-evolving is an elastic alternative of unsupervised learning, especially in the field of cloud computing anomaly detection. The self-evolving framework yields excellent cloud computing anomaly detection recipes \cite{guan2013adaptive,ji2016self,andonovski2018fault} and has prevailed in \textbf{these years}. Aside from being capable of replacing unsupervised detectors, self-evolved framework is also more appropriate for online learning.

Concurrent with our work, there are several efforts \cite{ruff2018deep,borghesi2018anomaly} to use neural network, especially deep learning for anomaly detection. In addition to being accurate, online learning would not be an issue as stochastic gradient descent (SGD) optimizer \cite{lecun1989backpropagation} were adopted in these methods. Although neural networks yield unusually brilliant results among computer vision and natural language processing, a critical result is that, it may not be suitable for cloud computing anomaly detection. The central limitations can be distilled to the following: (i) training a neural network requires expensive resources and cost plenty time (is time consuming), which is cost-inefficient; (ii) the cloud computing failure data is extreme imbalanced at which a regular neural network cannot handle; (iii) the cloud computing system log data can not meet the requirement of huge number of labeled instances for training a neural network.

Different from aforementioned work, our proposed framework pursuit a simple yet efficient realization for real-world cloud computing anomaly detection. We first customize a \textbf{serial} \textbf{or series} of models for class-agnostic and class-known anomaly detection respectively. The models can meet the requirements of classification, online learning and feature selection simultaneously. Our customized models can be considered as a combination of feature reduction task and an anomaly detection task (e.g. \cite{smith2010anomaly,fu2012hybrid}). Unlike PCA and ICA,which project original data into a new subspace and lose the original semantics of features \textbf{(semantic features}), we just evaluate each feature by weights. By leveraging the advantage of the model that can study incrementally, we can dynamically and automatically select the most important features for further study. Then we design a self-evolving framework that feeds predicted anomalies to\textbf{ training procedure to automatically adapt the model.} This design allows us to employ our work in a real-world application with an a better performance.

\section{Self-evolving Framework }
To build a unified architecture for cloud computing anomaly detection, we propose a holistically decoupled framework, leverage the self-evolving property to adjust our embedded customized detection model over iterations, and improve performance over time. Comparing with the existing two-stage detection methods (use PCA/ICA for dimension
reduction, then feed the proceeded data into a classifier), we designed several detectors for cloud computing anomaly detection in a more succinct reformulation. These detectors are homogeneous that simultaneously detect anomaly, learn model incrementally and select the most discriminatory attributes.

\subsection{Framework}
Table \ref{table:framework} summarize our proposed self-evolving anomaly detection framework on high level.

\begin{table}[!h]
\caption{A framework for proposed Self-evolving Anomaly Detection.}
\label{table:framework}
\begin{tabular}{l}
\hline
    \textbf{Input:} New Data $\mathbf{X}$.\\
    \textbf{Output:} Updated detector $\mathfrak{D}$.\\
    
    1. Initialize $\mathfrak{D}$. \\ 
    \textbf{While} $\mathbf{X}!=NULL$ \\
    \qquad 2. Predict label  $\mathbf{Y}=\mathfrak{D}$.\\
    \qquad 3. Find predicted negative labeled data $\mathbf{X}_{neg}$.\\
    \qquad 4. Send $\mathbf{X}_{neg}$ to Administrator and get true label $\mathbf{Y}_{neg}$. \\
    \qquad 5. Updated detector $\mathfrak{D}=update(\mathfrak{D},\mathbf{X}_{neg},\mathbf{Y}_{neg})$.\\
    \textbf{end while} \\ 
\hline
\end{tabular}
\end{table}

When our framework is first adopted on a new cloud computing environment, we selectively initialize the detector to identify most records as normal. The detector predicts records and sends predicted anomaly records to cloud operators for verification. These records are identified as either anomalies or normal states once a single epoch of cloud performance data record is available. Differing from other approaches that simply send all records to cloud operators, our method can greatly reduce the work of operators and achieve comparable result. All of these verified records would be used to update our detector. In our experience, our model will achieve a satisfactory result after ten epochs(300 examples each epoch). 

The pipeline of our architecture requires an embedded detector. Most current advances in cloud computing anomaly detection have been driven by combinations of dimension reduction methods and powerful basic statistical classifiers, such as  \cite{smith2010anomaly,huang2016related}. However, such a combination would not be suitable for our self-evolving framework since it is unable to adjust models in an incremental fashion.\textbf{ To solve that}, we propose OSEAD, which can detect anomalies with their abnormal reason. Moreover, by adding a sparsity regularization term, OSEAD can also select the most distinguishing attribute on records.

\subsection{General Anomaly Detection }
In this subsection, we introduce a novel incremental detector for cloud computing class-agnostic anomaly detection. 

Suppose we have a batch of $n$ data records, and each data record has $d$ attributes. For convenience, the batch of records can be denoted as $\mathbf{X}\in \mathbf{R}^{n\times d}$ in the real space. The label $y$ of a certain record is either 1 or -1, corresponding to a normal record or an anomaly record. Initially, we use a vector $\mathbf{Y}\in \mathbf{R}^{n}$ to represent the labels of $\mathbf{X}$.

For detection, we follow a classification plane fashion to build our models, similar to Least Squares Estimator (LSE) and Support Vector Machines (SVMs). Suppose the classification plane is $w$,  which is a column vector that $w \in \mathbf{R}^{d}$, our predictions of $\mathbf{X}$ can be denoted as $\mathbf{X}w$. Based on conventional statistical methods, we presented an incremental binary classifier, called L1LS, for our cloud computing anomaly detection. 

Least Squares Estimator \cite{friedman2001elements} is a profound machine learning method. It minimizes the residual sum of squared errors between true labels and predicted labels:
\begin{equation}
 \min \sum_{i=1}^{n}\left ( y_i-x_i*w-w_0 \right )^2
\end{equation}
where $w_{0}$ is the bias. For convenience,we rewrite the objective function by appending a column vector of 1 values and increasing the length of $w$ by 1 as following:
\begin{equation}
\label{formula:least_square}
 \min \left \| \mathbf{Y}-\mathbf{X}w \right \|_{2}^{2}
\end{equation}

Obviously, the classification plane $w$ can be considered as coefficients of attributes. For instance, $w_j$ is the weight of attribute $X^j$. The greater $w_j$ is, the more contribution attribute $X^j$ is made for detection. Hence, for selecting the most important attributes, we need to force most elements in $w$ close to 0. Inspired by related work on sparsity \cite{ng2004feature,mianjy2018stochastic}, we opt to modify (\ref{formula:least_square}) by adding a L1-norm regularization, thus the sparsity of $w$ is improved. The resulted formulation will be
\begin{equation}
\label{formula:L1LS}
 \min \left \| \mathbf{Y}-\mathbf{X}w \right \|_{2}^{2}+\lambda \left \| w \right \|_1
\end{equation}
where $\lambda$ is a trade-off factor between empirical loss and model sparsity. We call LSE with L1-regularization term as L1LS for short. 

However, L1LS can not be embedded in our self-evolving framework due to the following two shortcomings: the objective function of L1LS is not a convex function which leads to difficulty in solving, and it is not an incremental algorithm, which is a requirement of our framework. To meet these criteria, we employ the Stochastic gradient descent (SGD) method on our model. SGD only requires the first order derivative of our original model and can converge to a local optimal solution. Besides advantage conveniently solving objective, one big advantage of SGD is that it satisfy the requirement of incremental learning perfectly as SGD learns one example from anther example (or one batch examples from another batch examples). 

The derivative of \ref{formula:L1LS} is given as:
\begin{equation}
\label{formula:L1LS_derivative}
\frac{\partial \mathbb{F}}{\partial w} = -2\mathbf{X}'\left ( \mathbf{Y}-\mathbf{X}w \right )+\lambda sign\left ( w \right )
\end{equation}
where $sign\left ( \cdot  \right )$ denotes the element wise function that returns 1 if element value is greater than zero and -1 otherwise. Specially, if the element value is zero, we add a small value $1e^{-7}$ to avoid non-derivative. Empirically, we set the learning rate of SGD to $1e^{-4}$ for training \footnote{A dynamic selection manner is used to modulate the learning rate. For a quick training process, the initial learning rate is set to 0.1. Each time the objective value does not decrease, we divide learning rate by 10. The minimum value  of learning rate is $1e^{-7}$.}. After several iterations, our objective function would converge to a local minimum value and the sub-optimal $w$ would be obtained. The algorithm for solving L1LS is presented in Table \ref{table:L1LS_SGD}.
\begin{table}[h!]
\caption{Solving L1LS model using SGD.}
\label{table:L1LS_SGD}
\begin{tabular}{l}
\hline
    \textbf{Input:} new Data $\mathbf{X}$.\\
    \textbf{Output:} $w$.\\
    1. Initialize weight $w$, learning rate $lr$ and parameter $\lambda$. \\
    2. Compute the objective function value $\mathbb{F}$. \\
    3. Set $d = \infty$. \\
    
    \textbf{while} $d > 1e^{-6}$  \\
    \qquad 4. Compute derivative $der$ according to (\ref{formula:L1LS_derivative}); \\
    \qquad 5. Update objective function value $\mathbb{F}_{new} = \mathbb{F}_{previous} - lr*der$.\\
    \qquad 6. Update $d = \mathbb{F}_{previous} - \mathbb{F}_{new}$; \\
    \textbf{end while} \\ 
\hline
\end{tabular}
\end{table}

Once we have the output $w$, we are able to detect a cloud performance record $x$ by comparing the distance of $xw$ to 1 and to -1. If $xw$ is closer to 1, we will predict it as a normal record, otherwise it will be considered as an anomaly record.

\subsection{Detail Anomaly Detection}
In the previous subsection, we introduced a model named L1LS for detecting anomalies. L1LS is effective for detecting anomalies and is able to select the most important attributes for detection. However, it is a 2-class classifier that can only detect anomalies without their abnormal class. 

To generalize 2-class classifiers to multi-class classifiers, two paradigms are considered: "One vs Rest" and  "One vs One". "One vs Rest" strategy trains a single classifier per class. For $C$ class detection tasks, we need to train $C-1$ classifiers. Unlike "One vs Rest" strategy, "One vs One" strategy trains  $\frac{C\left ( C-1 \right )}{2}$ binary classifiers for a $C$-way multi-class problem. There is an obvious drawback to these two strategies: the training stages are expensive in both space and time.

Apart from the high time and space consumption, we argue that the above two strategies for multi-class classification problems would lose the competition between classes. They ignore the information that makes contribution to different classes. Instead of generalizing binary classifier to multi-class classifier, we develop a multi-class classifier based on L1LS by leveraging the benefit of one-hot encoder. 

Suppose $x$ is a performance record that belongs to $c$-th class, based on one-hot encoder, label $y$ would be denoted by a row zero value vector $y\in \mathbf{R}^{c}$ where $c$-th element value is one \footnote{In experiments, we found that using +1, -1 would achieve better results than using 0, +1. The phenomenon can be explained from two aspects: One is that our model is a distance based model instead of a possibility based model, the output is not in the range of $\left [0,1\right ]$. Also, the distance between -1 and +1 is greater than 0, +1. }. For example, if we have 5 classes and a record that belongs to $3$-th class, the label can be presented as $\left [ 0,0,1,0,0 \right ]$. For dataset $\mathbf{X} \in \mathbf{R}^{n*d}$, the corresponding labels $\mathbf{Y} \in \mathbf{R}^{n \times c}$ is a $n$ by $c$ matrix. 

\subsubsection{MCL1LS} 
Differing from L1LS of which the classification plane is a one-dimensional vector, the classification plane $\mathbf{W} \in \mathbf{R}^{d \times c}$ of our new multi-class L1-norm Least Squares method (MCL1LS in short) is a c-dimensional matrix. A general MCL1LS model can be written as:
\begin{equation}
\label{formula:MCL1LS}
    \min \left \| \mathbf{Y}-\mathbf{XW} \right \|_{2}^{2}+\lambda \left \| \mathbf{W} \right \|_1
\end{equation}
 and the related derivative can be written as follows:
\begin{equation}
\frac{\partial \mathbb{F}}{\partial \mathbf{W}} = -2\mathbf{X}'\left ( \mathbf{Y}-\mathbf{XW} \right )+\lambda sign\left ( \mathbf{W} \right )
\end{equation}

Although \ref{formula:MCL1LS} shares a similar formulation with \ref{formula:L1LS}, it is noteworthy that $\mathbf{W}$ in MCL1LS is a matrix while $w$ in L1LS is a column vector. Such a modification enables L1LS to detect multi-class records. 

Once we obtain the desired $\mathbf{W}$, the prediction of a record $x$ can be presented by $x\mathbf{W}$ and the index of the max value in vector $x\mathbf{W}$ indicate the predicted class of $x$.

\subsubsection{MCL21LS} MCL1LS achieve a significant improvements in class-known anomaly detection as demonstrated in previous subsection. However, it is not able to select the most distinguishing attributes. For attribute selection, it is required that most rows in $\mathbf{W}$ are close to zero instead of elements. An L1-norm based regularization term, however, is not satisfied to the matrix as it can not deal with row sparsity. In order to introduce row sparsity to $\mathbf{W}$, an alternative model based on L21-norm regularization term is provided.

The use of L21-norm regularization terms back to $R_1-PCA$ \cite{ding2006r}, if not earlier. Differ from the sparsity requirement of L1-norm regularization term, L21-norm regularization term also requires the row sparsity. For a matrix $\mathbf{W} \in \mathbf{R}^{d \times c}$,  L21-norm is defined as:
\begin{equation}
\label{formula:L21Norm}
    \left \|  \mathbf{W}\right \|_{2,1}=\sum_{i=1}^{d}\sqrt{\sum_{j=1}^{c}w_{ij}^2}=\sum_{i=1}^n\left \| w^i \right \|_2
\end{equation}
where $w_{ij}$ means the $i$-th row $j$-th column element in $\mathbf{W}$ and $w^i$ denotes the $i$-th row vector in $\mathbf{W}$.

By adapting L21-norm regularization term, the problem of MCL1LS changes to:
\begin{equation}
\label{formula:MCL21LS}
    \min \left \| \mathbf{Y}-\mathbf{XW} \right \|_{2}^{2}+\lambda \left \| \mathbf{W} \right \|_{2,1}
\end{equation}

For simplicity, we call this model as Multi-Class L21-norm Least Squares Method (MCL21LS in short). Moreover, MCL21LS is more effective in selecting the distinguishing attributes.

Although Nie et al. \cite{nie2010efficient} presented a simple algorithm to solve the L21-norm problem, we still follow the solution in Table \ref{table:L1LS_SGD} as it is compatible with online learning. According to Eq.\ref{formula:L21Norm}, we give the derivative of L21-norm $\mathbf{W}$ as follows:
\begin{align}
\frac{\partial\left \| \mathbf{W} \right \|_{2,1}}{\partial \mathbf{W}} 
& = \left [ \frac{\partial\left ( \sum_{i=1}^{d}\left \| w_i \right \|_2 \right )}{\partial w_j} \right ]_{d \times 1} \\
& = \left [ \frac{\partial\left (     \sum_{i=1}^{d}\left ( w_iw_i^T \right )^{\frac{1}{2}} \right )}{\partial w_j} \right ]_{d \times 1} \\
& = \left [ \frac{w_j}{\left \| w_j \right \|_2} \right ]_{d \times 1} \\
& = \begin{bmatrix}
\frac{1}{\left \| w_1 \right \|_2} &  &  & \\ 
 &  \frac{1}{\left \| w_2 \right \|_2}&  & \\ 
 &  &  \ddots & \\ 
 &  &  & \frac{1}{\left \| w_d \right \|_2}
\end{bmatrix}
\begin{bmatrix}
w_1\\ 
w_2\\ 
\vdots \\ 
w_d
\end{bmatrix}\\
& = \begin{bmatrix}
\frac{1}{\left \| w_1 \right \|_2} &  &  & \\ 
 &  \frac{1}{\left \| w_2 \right \|_2}&  & \\ 
 &  &  \ddots & \\ 
 &  &  & \frac{1}{\left \| w_d \right \|_2}
\end{bmatrix}\mathbf{W}\\
& = \Sigma \mathbf{W}
\end{align}
where $\Sigma$ is a diagonal matrix. Thus, the derivative of \ref{formula:MCL21LS} w.r.t $W$ can be presented as:
\begin{equation}
\frac{\partial \mathbb{F}}{\partial \mathbf{W}}=-2\mathbf{X}^T(\mathbf{Y}-\mathbf{XW})+\lambda \Sigma \mathbf{W}
\end{equation}

Based on Table \ref{table:L1LS_SGD}, we can solve problem \ref{formula:MCL21LS} efficiently.

\section{Experiments}
In this section, we present comprehensive experiments to demonstrate the effectiveness of our proposed framework and the embedded detection methods for cloud computing anomaly detection. The MATLAB codes of our proposed frameworks and implement details are made available at \hyperref[https://github.com/13952522076/incrementalAD]{https://github.com/13952522076/incrementalAD}.

\subsection{Data Sets}
We collected performance records from a cloud computing environment that consists of 362 servers and connected by gigabit Ethernet. The cloud servers are equipped with two to four Intel Xeon or AMD Opteron cores and 2.5 to 8 GB of RAM. We have installed Xen 3.1.2 hypervisors on the cloud servers. The operating system on a virtual machine is Linux 2.6.18 as distributed with Xen 3.1.2. Each cloud server hosts up to eight VMs. A VM is assigned up to two VCPUs, among which the number of active ones depends on applications. The amount of memory allocated to a VM is set to 512 MB. We run the RUBiS \cite{cecchet2002performance} distributed online service benchmark and MapReduce \cite{dean2008mapreduce} jobs as cloud applications on VMs. The applications are submitted to the cloud computing system through a web based interface. We have also developed a fault injection program, which is able to randomly inject four major types with 17 sub-types of faults to cloud servers. They mimic faults from CPU, memory, disk, and network. In order to ease the reproduction of our work and promote the development of related research, the could computing performance data set is made publicly available\footnote{http://dcslab.cse.unt.edu/~song/Data/}. The data is normalized to the range $[0,1]$ with respect to attributes before detection.

\subsection{Main result}
We empirically evaluate our proposed self-evolving framework and our customized detectors by sensitivity and specificity. Besides, other metrics like accuracy and F1-measure value are considered. Since our detectors are all incremental learning methods, we set the epoch size to 300 in all our experiments unless otherwise specified. It is worth noticing that when detecting detailed anomaly types, records in each anomaly type is extremely rare. If the imbalance problem is not alleviated in training, it introduces large, disastrous bias to normal records, and training ignores anomalies. As a simple solution, we adopt an over-sampling technique, named SMOTE \cite{chawla2002smote}, to over sample anomalies and re-balance the data distribution. The k-nearest neighbours parameter. Because of time-consuming constraints, the generated data amount is set same as original data amount. 

Besides advantage high performance, one big advantage of our self-evolving method is that we only requires labels of partial data and consequently reduce the work of administrator. However, its capacity to reducing labeling requirement is not fully explored. To further investigate the property, we initialize weights that lead the predictions bias towards normal records as they are dominate. Instead of randomly initialize weights that follows Gaussian distribution, we initialize $w=rand()+\alpha X^{-1}Xe$, where $\alpha$ is a parameter that regulate anomaly rate and $rand()$ is a function that generate weights follow standard Gaussian distribution. The reduction of labeling work will be revealed latter.

\subsubsection{L1LS study}

\begin{figure}[!h]
    \centering
    \includegraphics[width=0.9\linewidth]{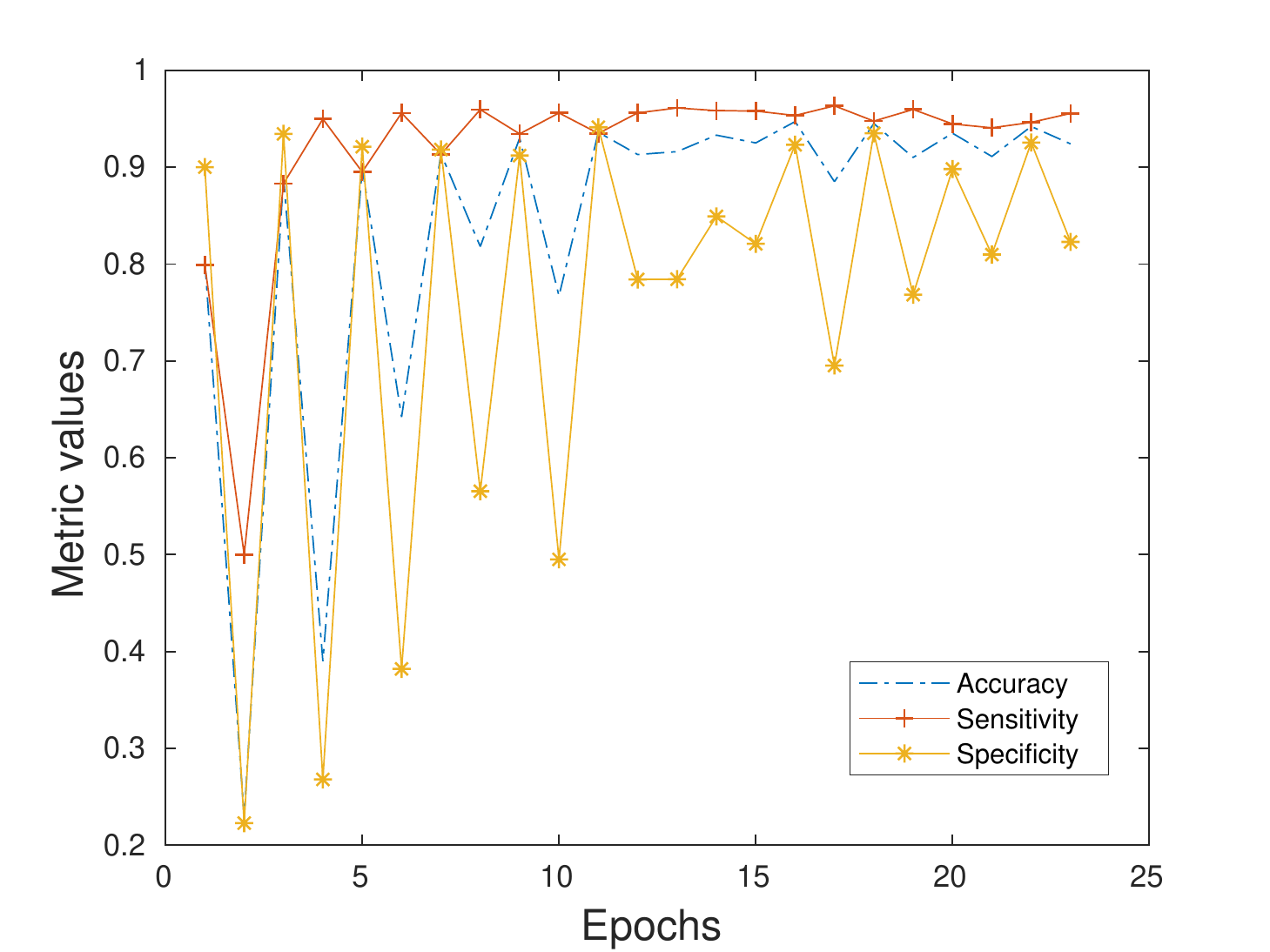}
    \caption{Performance of self-evolving framework with L1LS. }
    \label{fig:1_L1LS}
\end{figure}

Fig. \ref{fig:1_L1LS} depicts the performances of our L1LS embedded self-evolving framework. It is worth noticing that the specificity fluctuates drastically in the early epochs and then converges to the range of 82\% and 92\%. Similar phenomenon  is also shown on the accuracy, suggesting that our system is effective and converges quickly in about 10 epochs. Somewhat surprisingly, the sensitivity always leads to a high value (around 95\%). This may be due to our weights initialization method and the fact that there is more normal records in the data set.

\subsubsection{MCL21LS study}
\begin{figure*}[!t]
    \centering
    \subfloat[]{
    \includegraphics[width=0.15\linewidth]{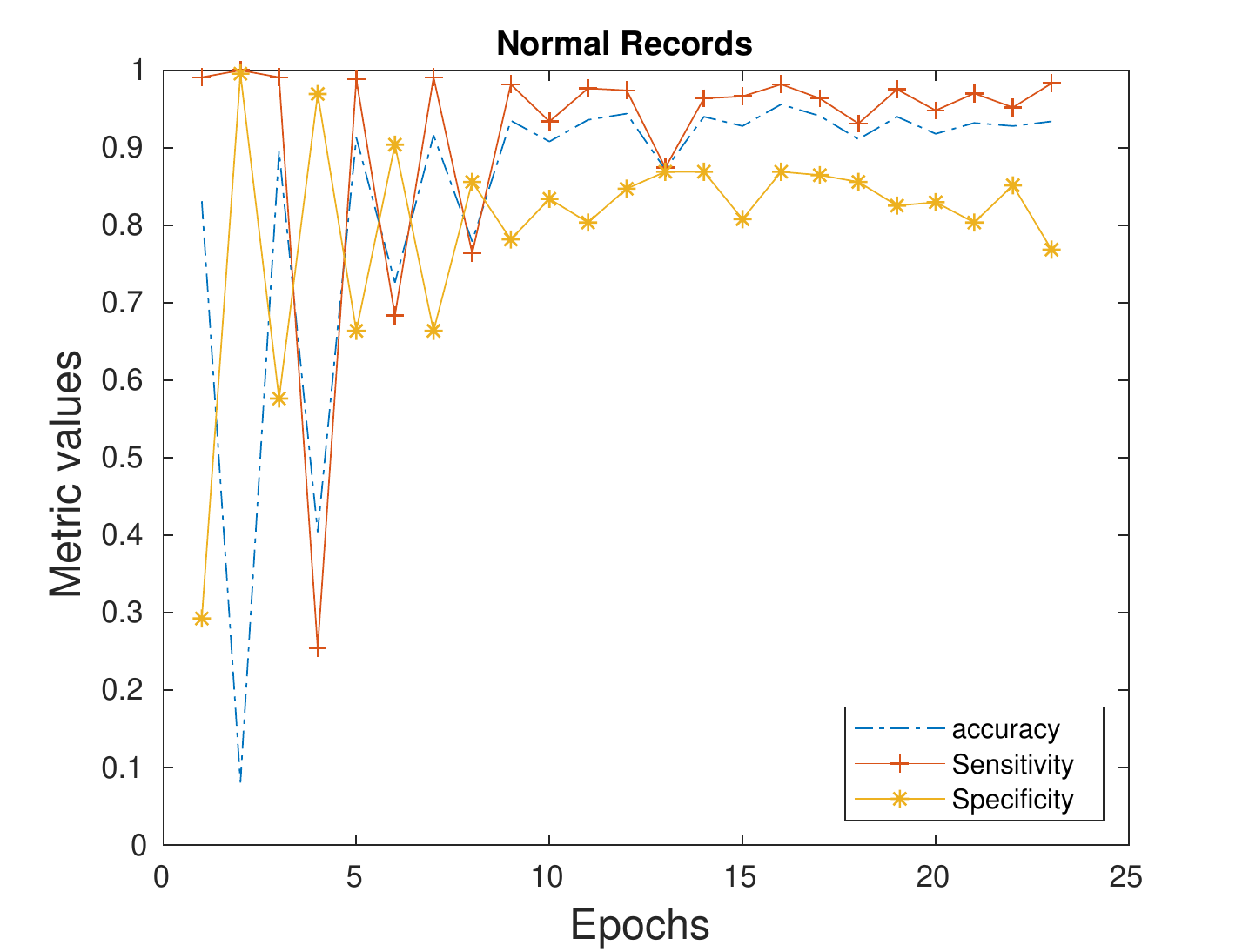}
    }
    \subfloat[]{
    \includegraphics[width=0.15\linewidth]{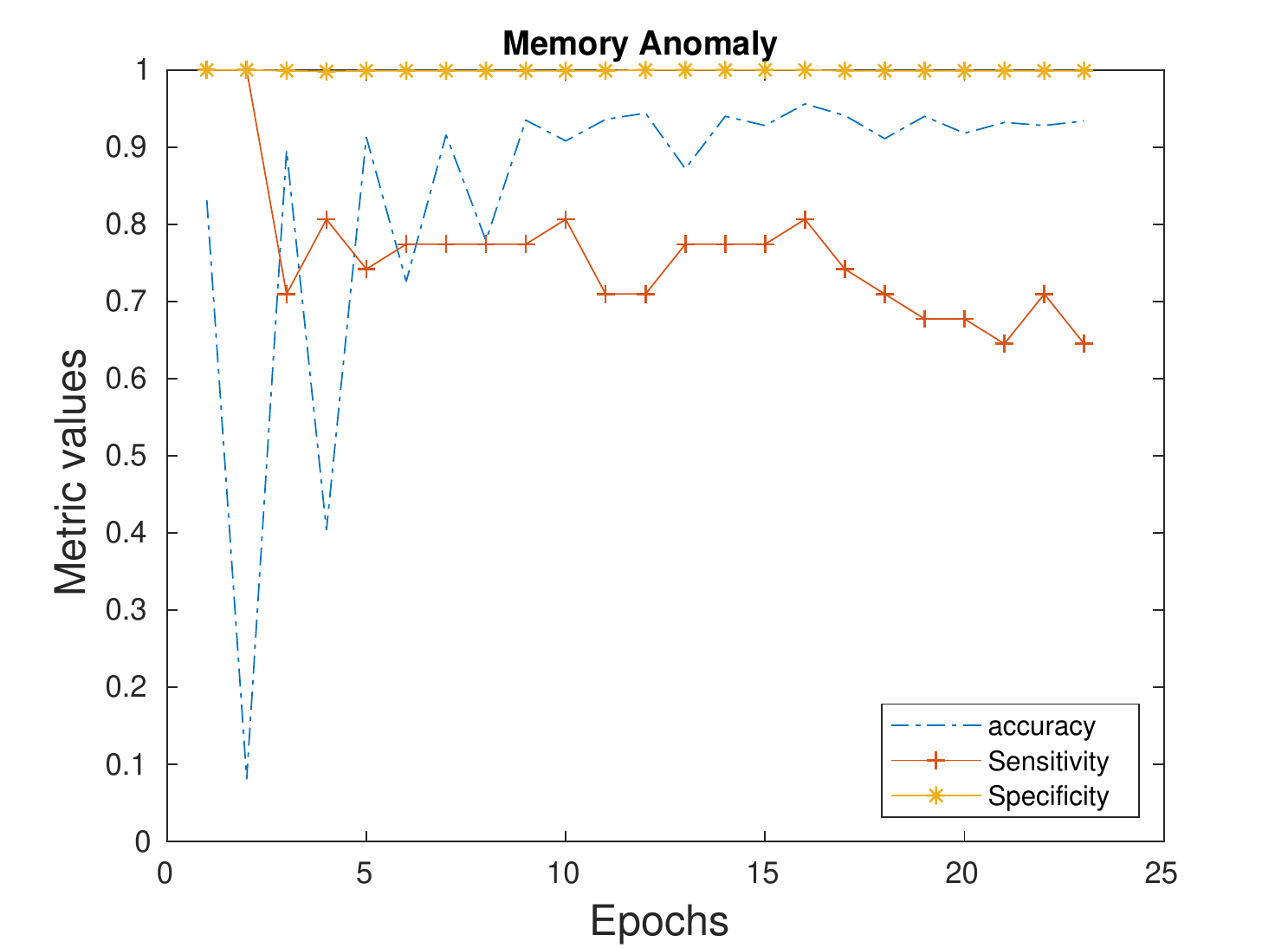}
    }
    \subfloat[]{
    \includegraphics[width=0.15\linewidth]{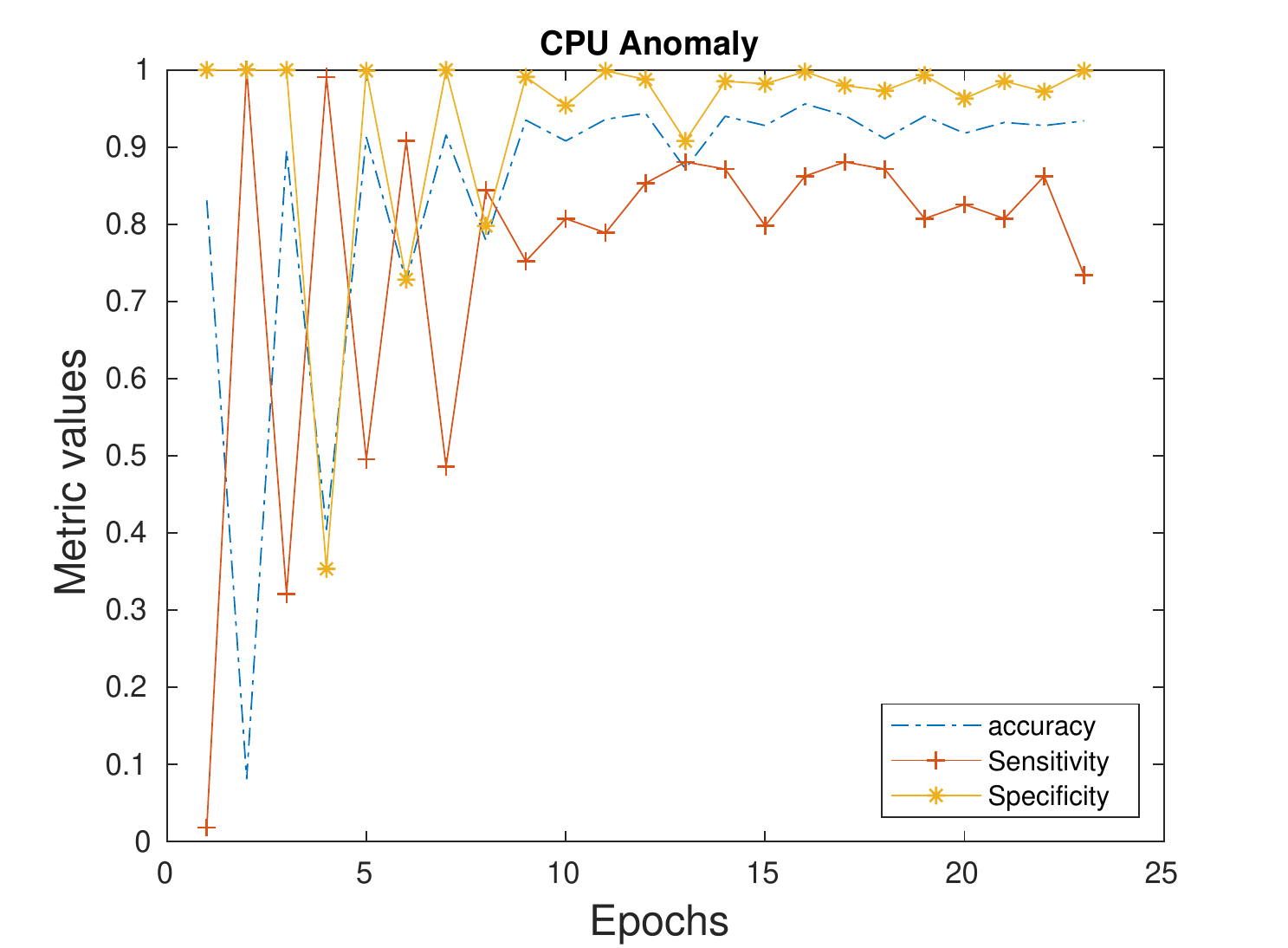}
    }
    \subfloat[]{
    \includegraphics[width=0.15\linewidth]{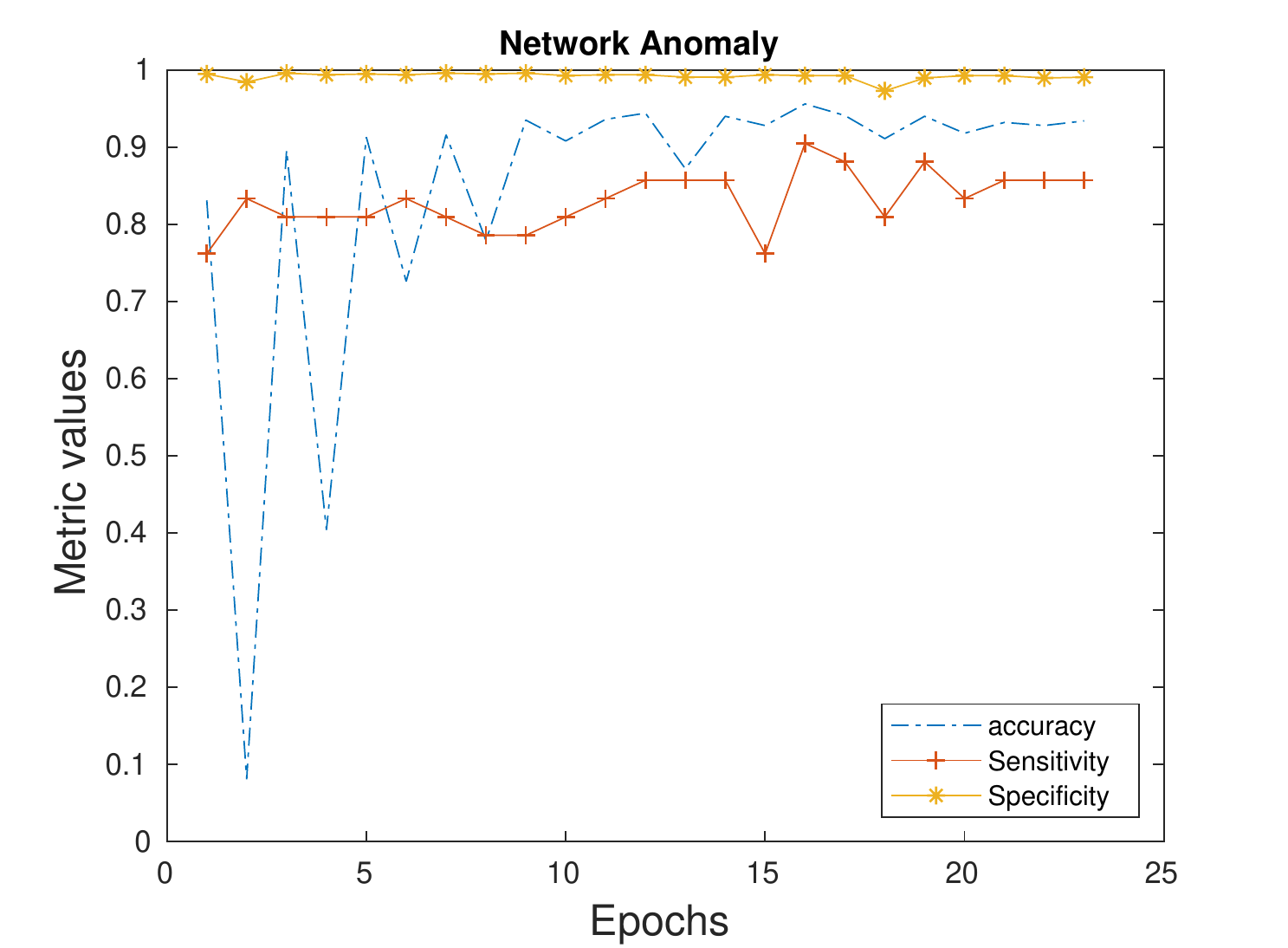}
    }
    \subfloat[]{
    \includegraphics[width=0.15\linewidth]{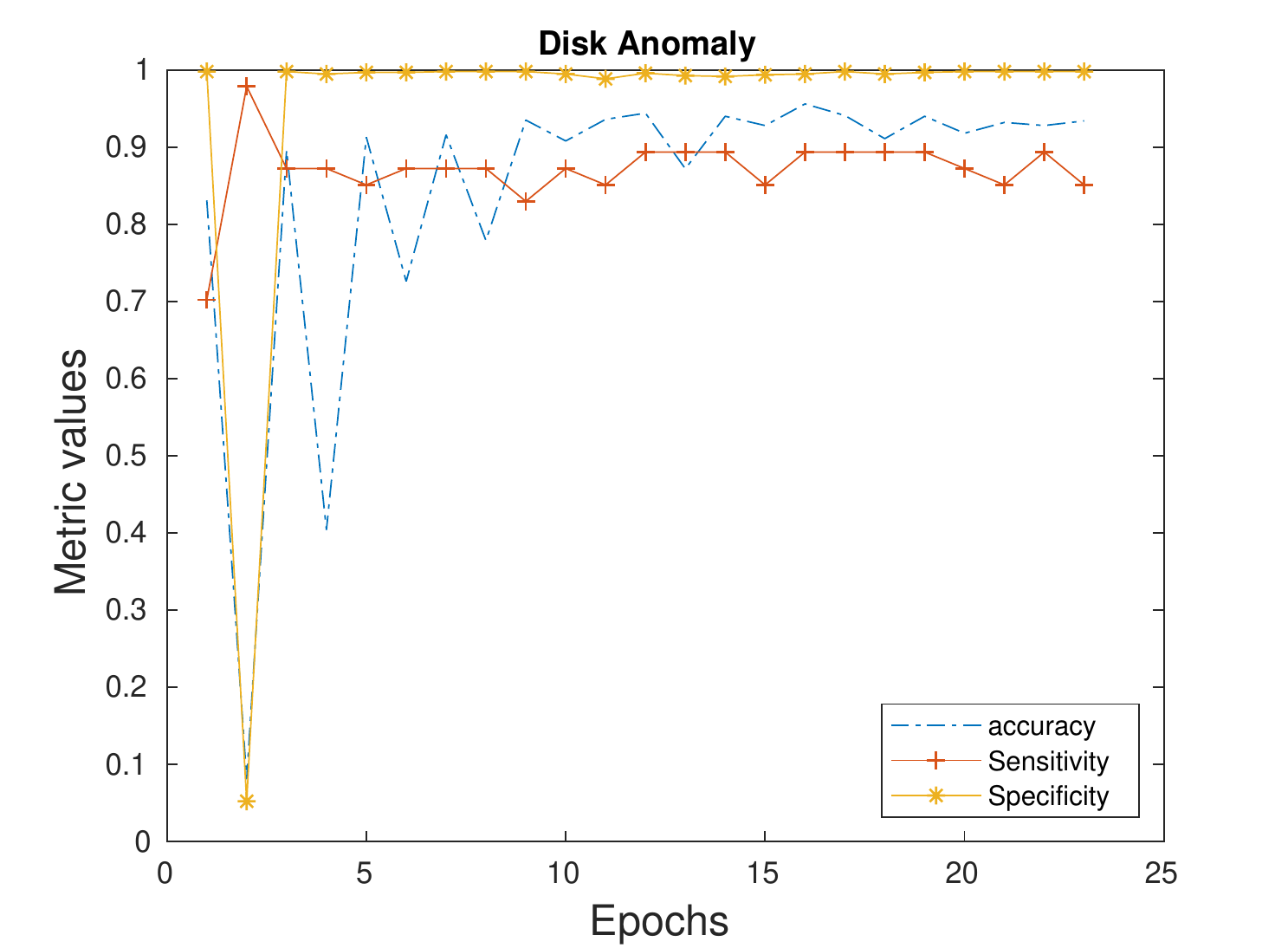}
    }
    \caption{Performance of self-evolving framework with MCL21LS. From left to right are detection performances on normal records, Memory anomaly, CPU anomaly, Network Anomaly and Disk anomaly.}
    \label{fig:1_MCL21LS}
\end{figure*}

Our self-evolving framework with MCL21LS is able to identify detail anomaly types of performance records in the cloud performance data. Anomalies in our data set are composed of memory anomalies, CPU anomalies, network anomalies and disk anomalies. Fig. \ref{fig:1_MCL21LS} shows the detection performance of our self-evolving framework with MCL21LS on each data type respectively.

Similar to \cite{fu2012hybrid}, we reported sensitivity, specificity and additional accuracy in figures. Broadly speaking, our system performs better and better on all record types as the epoch increases. After about ten epochs, detection performance gets stabilizes and almost does not increase or decrease further with negligible fluctuations. The mean sensitivity of last five epochs over all five record types (including normal records and anomaly records) achieves 83.47\%, and the corresponding mean specificity achieves 95.72\%. Most remarkably, our self-evolving framework achieves fairly high specificity on all anomaly types, which indicates that almost all our predicted anomalies are true anomalies.  

\begin{figure}[!htbp]
    \centering
    \subfloat[]{
    \includegraphics[width=0.3\linewidth]{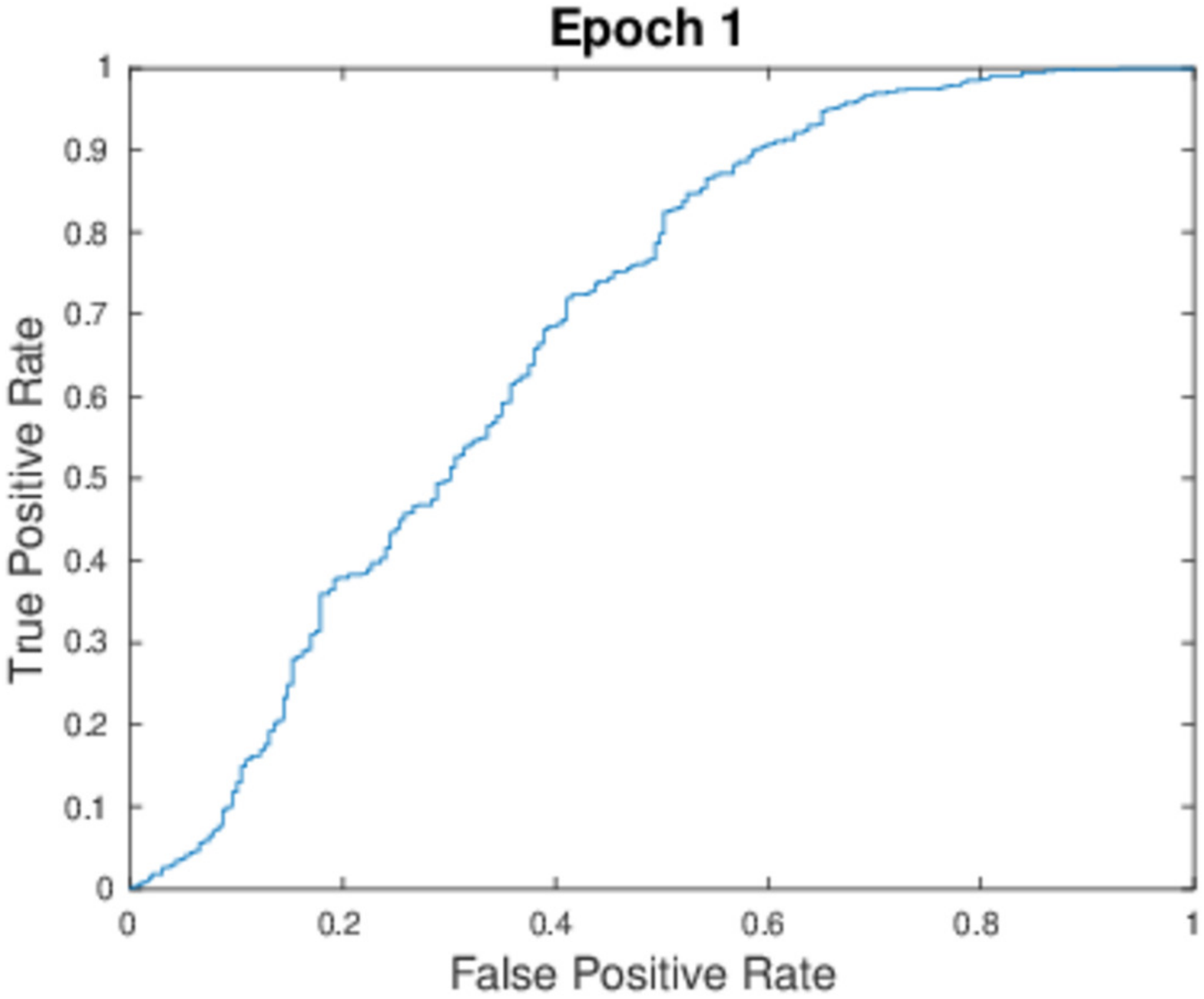}
    }
    \subfloat[]{
    \includegraphics[width=0.3\linewidth]{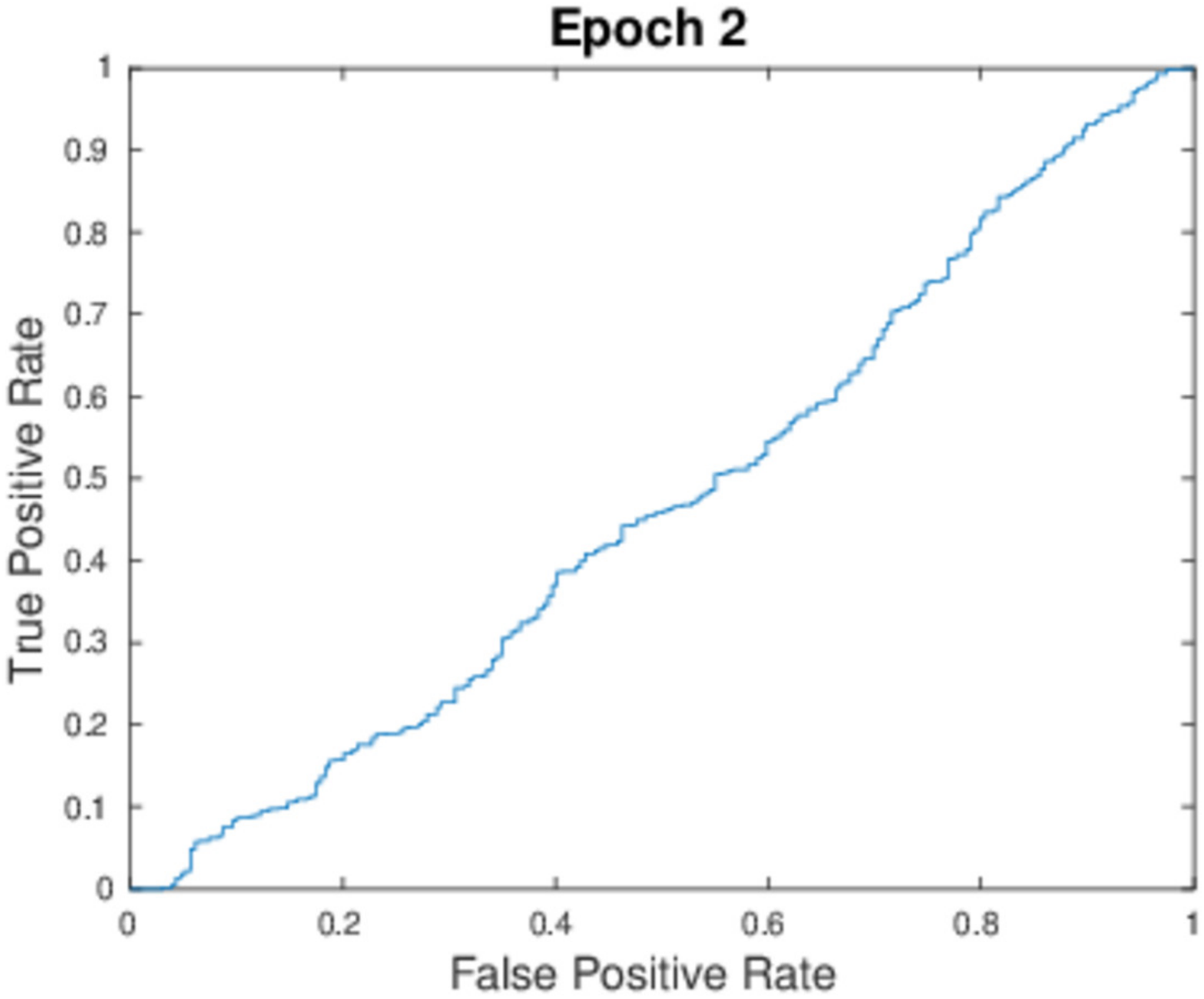}
    }
    \subfloat[]{
    \includegraphics[width=0.3\linewidth]{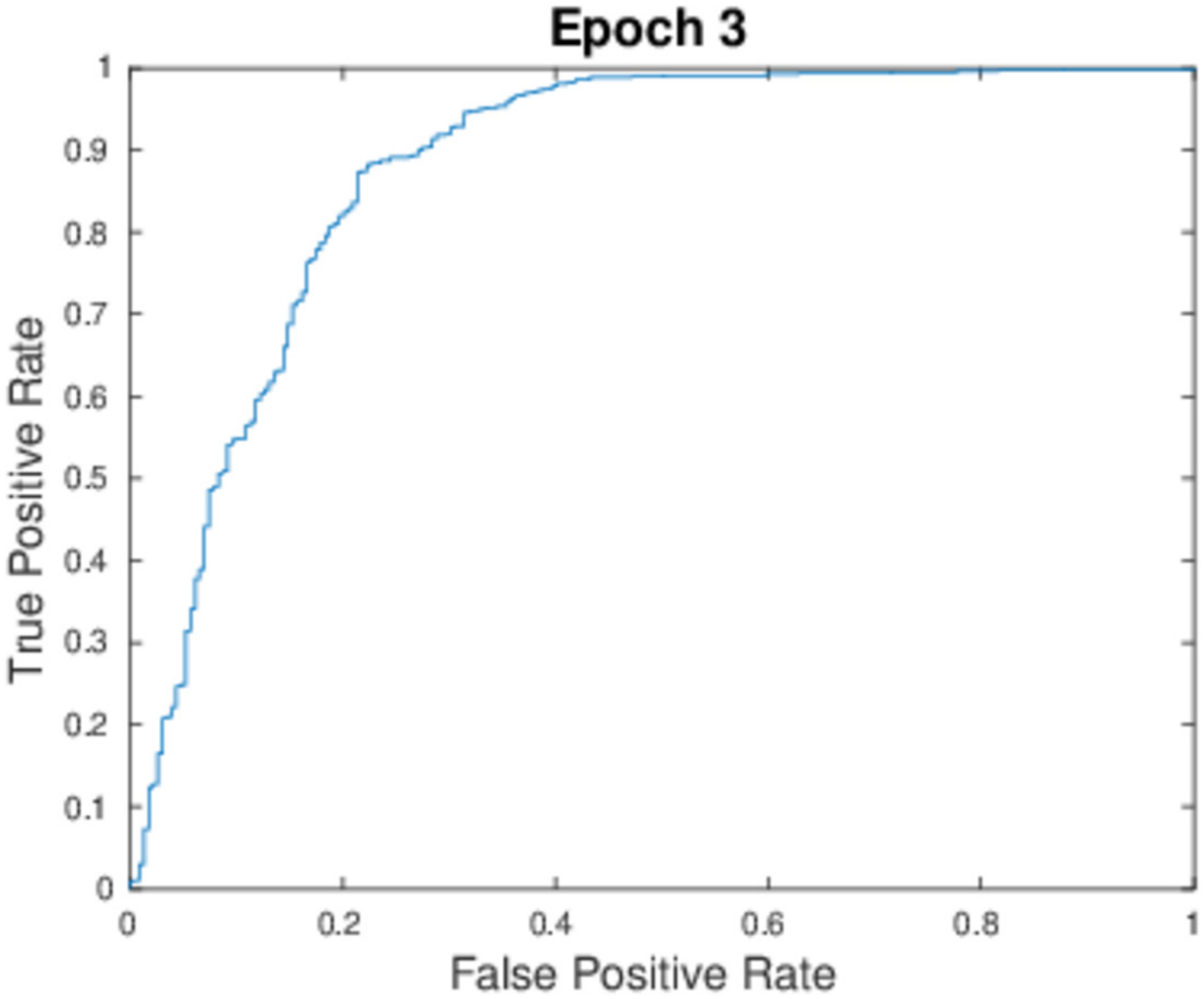}
    }
    \\
    \subfloat[]{
    \includegraphics[width=0.3\linewidth]{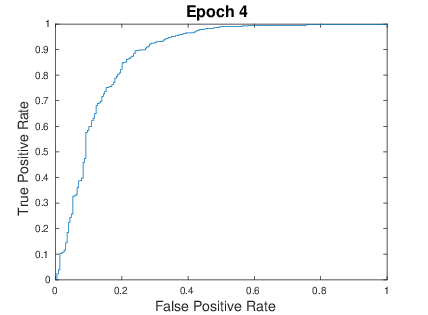}
    }
    \subfloat[]{
    \includegraphics[width=0.3\linewidth]{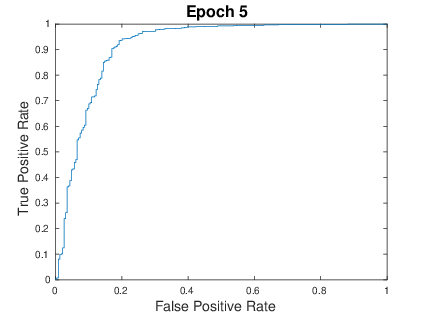}
    }
    \subfloat[]{
    \includegraphics[width=0.3\linewidth]{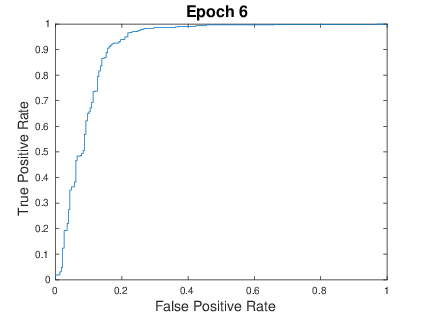}
    }
    \\
    \subfloat[]{
    \includegraphics[width=0.3\linewidth]{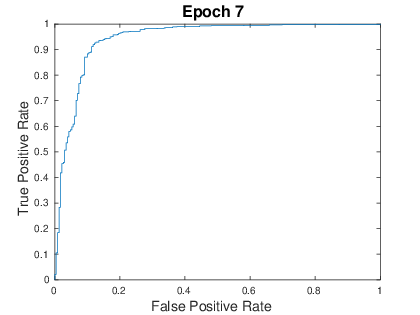}
    }
    \subfloat[]{
    \includegraphics[width=0.3\linewidth]{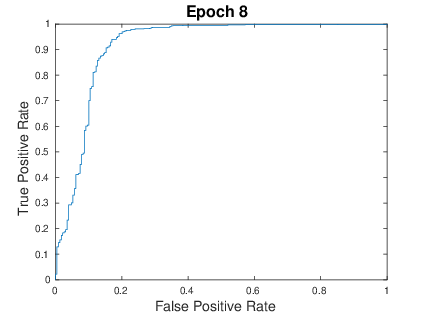}
    }
    \subfloat[]{
    \includegraphics[width=0.3\linewidth]{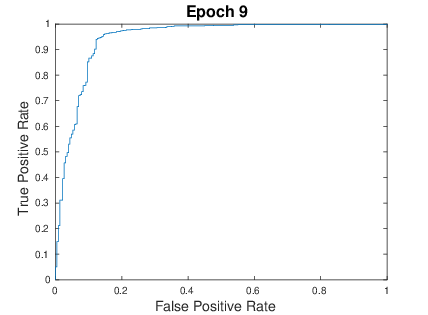}
    }
    \\
    \subfloat[]{
    \includegraphics[width=0.3\linewidth]{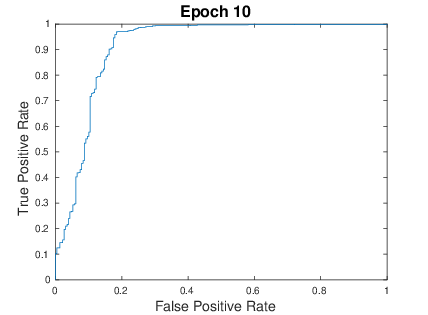}
    }
    \subfloat[]{
    \includegraphics[width=0.3\linewidth]{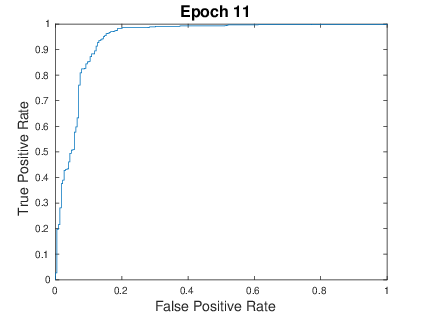}
    }
    \subfloat[]{
    \includegraphics[width=0.3\linewidth]{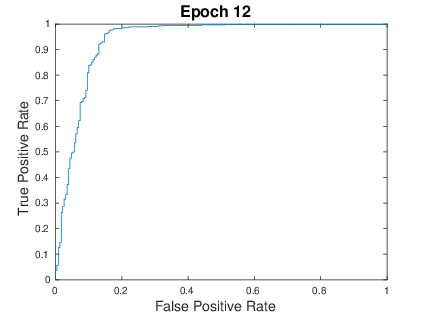}
    }
    \\
    \subfloat[]{
    \includegraphics[width=0.305\linewidth]{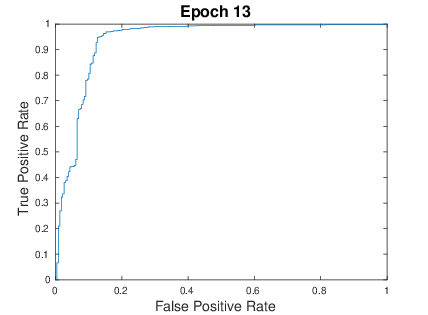}
    }
   \subfloat[]{
    \includegraphics[width=0.305\linewidth]{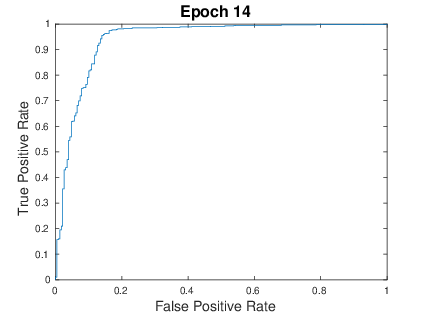}
    }
    \subfloat[]{
    \includegraphics[width=0.305\linewidth]{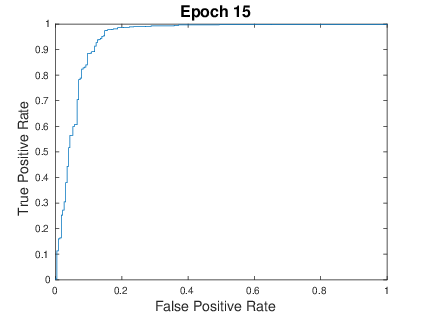}
    }
    \caption{ROC curves of first 15 epochs.}
    \label{fig:ROC}
\end{figure}

\begin{figure}[!htbp]
    \centering
    \includegraphics[width=0.9\linewidth]{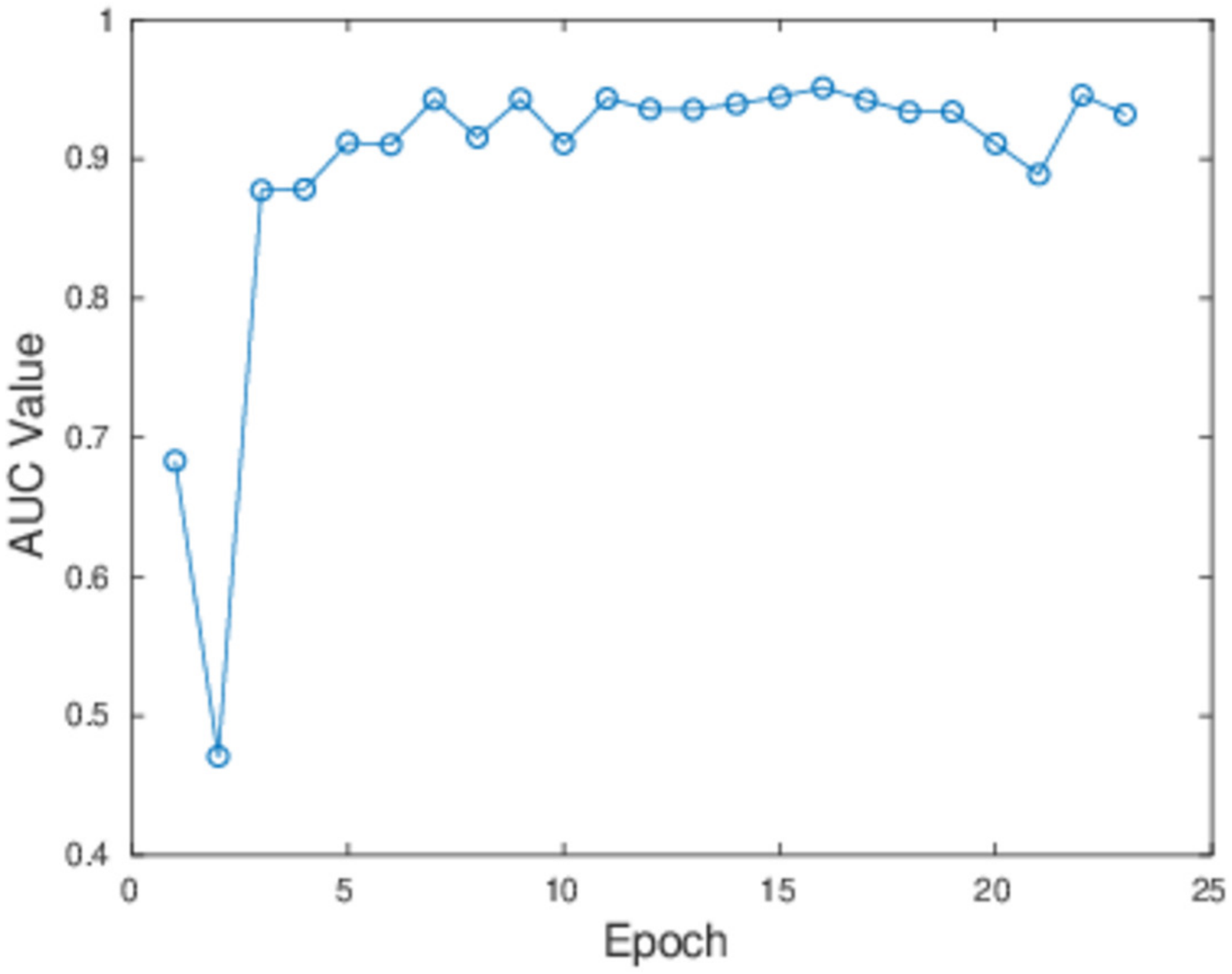}
    \caption{AUC values of each epoch.}
    \label{fig:AUC}
\end{figure}

Moreover, we report the ROC curve of each epoch since ROC curve is independent of the data distribution and could better demonstrate the performance of detectors. Due to length limit, we plot first 15 epochs ROC curves in Fig.\ref{fig:ROC} and all the AUC values are plot in Fig.\ref{fig:AUC}. As can been seen in Fig. \ref{fig:ROC} and Fig. \ref{fig:AUC}, our system performs better and better as epochs increase and stabilizes then and the AUC will stabilizes at 0.94. Such a high AUC value explicitly demonstrates that our system would not be influenced by the imbalance data distribution. 

Across all aforementioned results we observe that 1) our MCL21LS based self-evolving framework will progressively achieve better performance as more data are available and get stabilized in a short time; 2)  our framework is able to detect various types of anomalies simultaneously and precisely; 3) regardless of any metric used to evaluate our approach, our approach performs very well intuitively. As demonstrated in Fig. \ref{fig:1_L1LS}and \ref{fig:1_MCL21LS}, our approach is effective and practical in real applications.

\subsubsection{Comparison}

We have also compared our algorithm with anomaly detectors using some advanced learning algorithms,including the ensemble of Bayesian sub-models and decision tree classifiers \cite{guan2012ensemble}, hybrid 1 and 2-class SVM \cite{fu2012hybrid}. We showcase the detail performance of various methods in Table \ref{table:Comparison}. Since our approach is an online learning fashion, we report the last five epochs average performance for comparison.
\begin{table}[]
    \centering
    \caption{Comparison with state-of-art methods.}
    \begin{tabular}[width=1\linewidth]{c|cc}
    \hline
     & Sensitivity & Specificity \\ \hline
    Ensemble \cite{guan2012ensemble} &72.5\% &- \\
    Hybrid  \cite{fu2012hybrid}&92.1\% &83.8\% \\ \hline
    Ours(L1LS) &\textbf{94.90\%} &85.99\% \\
    Ours(MCL21LS) & 83.47\% & \textbf{95.72}\%\\\hline
    \end{tabular}
    \label{table:Comparison}
\end{table}

In Table \ref{table:Comparison} we compare with previous state-of-art methods. We find our framework with L1LS achieves the best sensitivity at 94.90\% but not best specificity, while L21LS achieves the best specificity at 95.72\% but not best sensitivity. The key insight behind this phenomenon is that L1LS is especially customized for type-agnostic anomaly detection while L21LS is customized for type-known anomaly detection. Aside from better performance, one big advantage is that our self-evolving framework is able to modulate detector incrementally, which is not supported by others. These two characteristics make our approach is practical for real-world could computing system analysis. Furthermore, our approach is able to adaptively select the most distinguishing attributes for further study. It will be explored later.

\subsubsection{Labeling Efficiency} One distinct advantage of our self-evolving framework is the less labeling requirement. Different from previous work that require labels of all records, our approach only require administrator to label the predicted abnormal records in each epoch, which drastically reduce the labeling work in real applications. We argue that most of the records make no contributes to training detectors and empirically show the performance of self-evolving framework is comparable with traditional counterpart. 
\begin{figure}[!htb]
    \centering
    \includegraphics[width=1.1\linewidth]{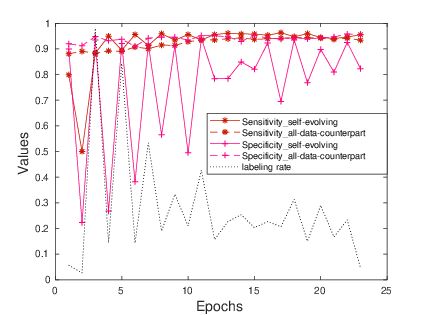}
    \caption{Performance of self-evolving framework VS. all-evolving counterpart.}
    \label{fig:label}
\end{figure}

From Fig. \ref{fig:label} we have the following interesting observations. First, our self-evolving framework achieves almost the same sensitivity as all-evolving counterpart, which demonstrate that our self-evolving framework is sensitive to the normal records. Second, although we achieve the comparable sensitivity, the specificity of our self-evolving framework is mediocre, the insufficient learning leads a little drop of specificity compares with all-evolving counterpart. However, such a degradation is acceptable if we take the contribution of labeling reduction into account. As shown in Fig. \ref{fig:label}, we only need to label 21.06\% (mean value of last 10 epochs) of the original data to get satisfactory results.
This important observation clearly demonstrates the efficiency of our self-evolving framework and empirically justifies our motivation to use part record to incrementally adjust our customized detectors.

\subsection{Ablation Study}
In order to understand the effectiveness of our proposed approach, we run models with different settings on provided data set and record their evaluations in Table \ref{table:ablation}.

\begin{table*}[!htbp]
    \centering
    \caption{Ablation study : Effects of various settings of our proposed frameworks.}
    \begin{tabular}{l|cccccccc|cccccccc}
    \hline
     & \multicolumn{16}{c}{Self-evolving framework (epoch size:300)} \\ \cline{2-17} 
     & \multicolumn{8}{c|}{L1LS} & \multicolumn{8}{c}{MCL21LS} \\ \hline
    SMOTE & & \checkmark & & & \checkmark & & \checkmark & \checkmark & & \checkmark & & & \checkmark & & \checkmark & \checkmark  \\
    Biased Init & & & \checkmark & & \checkmark & \checkmark & & \checkmark & & & \checkmark & & \checkmark & \checkmark & & \checkmark  \\
    Self-evolving & & & & \checkmark & & \checkmark & \checkmark & \checkmark &  & & & \checkmark & & \checkmark & \checkmark & \checkmark\\ \hline
    Sensitivity & 93.70& 95.38& 93.98& 95.05& 95.47& 94.93& 95.71& 96.06& 
    84.81& 87.49& 84.81& 84.01& 87.49& 83.47& 88.10 &88.94\\
    Specificity & 94.79 & 94.05& 94.86& 76.38& 94.07& 84.49 & 72.99 & 76.11& 
    96.16& 96.89& 96.16& 96.07& 96.89& 95.72& 96.01 &96.40\\\hline
    \end{tabular}
    \label{table:ablation}
    
\end{table*}

By adding SMOTE data up-sampling method, we can see the sensitivity is improving both on L1LS and MCL21LS, and we also achieve higher sensitivity compare with counterparts without SMOTE. We do not see much difference in specificity.

When adding the biased initialization module, the performance shows negligible difference, indicating that the biased initialization module have no influence on anomaly detection. This is consistent with our expectations, because the main idea of this module is to reduce the labeling work.

As an important module in this paper, we analysis the impact of self-evolving framework. As the labeling work efficiency is investigated previous, here we reveal its impact on performance. For L1LS based framework, we see that there always a decreased specificity if we add the self-evolving module. On the contrary, our MCL21LS based self-evolving framework achieves a high specificity constantly. An interesting observation is that sensitivity values obtained by L1LS is often higher than specificity values, and MCL21LS is the opposite. A potential reason is that L1LS based framework is designed for type-agnostic anomaly detection while MCL21LS counterpart is specifically designed for type-known anomaly detection, which can detect detail types of anomalies.

\subsection{Attribute Selection}
Next, we disentangle the detector's influence on attribute selection. For this purpose, we investigate formulations of L1LS and MCL21LS. As shown in \ref{formula:L1LS},\ref{formula:MCL1LS} and \ref{formula:MCL21LS}, all the objective functions are composed of two modules. The first module is loss function that aims to fitting the distribution of input data, and the second module is the regularization term that avoid over-fitting and improve model generalization.  As a trick, we use L1/L21-norm based regularization terms instead of conventional squared L2-norm based ones, which additionally provide sparsity\cite{cai2018feature,wen2019robust} for pursuant weights and make our models possible to select the most distinguishing attributes. To evaluate the importance of each attributes, we rank the attributes by the L2-norm distance of corresponding rows in weights matrix. Experimentally, we tun parameter $\lambda$ in the range of $[0.01, 0.1, 1, 10]$. The top ten ranked attributes are selected.

\begin{table*}[]
\centering
\caption{Attributes selection on L1LS based Self-Evolving Framework}
\label{tab:selection_L1LS}
\begin{tabular}[width=0.9\linewidth]{c|cc|cc|cc|cc}
\hline
\multirow{2}{*}{rank} &
\multicolumn{2}{c|}{$\lambda = 0.01$} &
\multicolumn{2}{c|}{$\lambda = 0.1$} & 
\multicolumn{2}{c|}{$\lambda = 1$} & 
\multicolumn{2}{c}{$\lambda = 10$} \\ \cline{2-9}
& weights & Attributes & weights & Attributes & weights & Attributes & weights & Attributes \\\hline
\#1& 1.1563 & runq-sz & 1.1378 & runq-sz & 0.6223 & runq-sz & 0.0337 & oseg/s \\
\#2& 0.8523 & plist-sz & 0.7171 & ldavg-1 & 0.3141 & plist-sz & 0.0322 & \%vmeff \\
\#3& 0.7992 & ldavg-1 & 0.7068 & plist-sz & 0.2975 & cswch/s & 0.0318 & \%soft 4 \\
\#4& 0.6899 & tcpsck & 0.5946 & tcpsck & 0.2794 & tcpsck & 0.0267 & idel/s \\
\#5& 0.5683 & cswch/s & 0.564 & cswch/s & 0.2744 & ldavg-1 & 0.0261 & txpck/s eth0 \\
\#6& 0.5178 & bufpg/s & 0.4248 & totsck & 0.2306 & udp6sck & 0.026 & orq/s \\
\#7& 0.4177 & svctm dev8-0 & 0.3811 & bufpg/s & 0.1696 & \%util dev253-1 & 0.0238 & pgsteal/s \\
\#8& 0.3991 & totsck & 0.3382 & \%soft 1 & 0.1357 & totsck & 0.0222 & \%usr 4 \\
\#9& 0.3556 & \%soft 1 & 0.3321 & svctm dev8-0 & 0.1194 & pgscank/s & 0.0221 & tps \\
\#10& 0.3528 & kbcached & 0.2987 & ldavg-15 & 0.1112 & \%usr 4 & 0.0215 & \%util dev253-0 \\\hline
\end{tabular}
\end{table*}

\begin{table*}[]
\centering
\caption{Attributes selection on MCL21LS based Self-Evolving Framework}
\label{tab:selection_MCL21LS}
\begin{tabular}[width=0.9\linewidth]{c|cc|cc|cc|cc}
\hline
\multirow{2}{*}{rank} &
\multicolumn{2}{c|}{$\lambda = 0.01$} &
\multicolumn{2}{c|}{$\lambda = 0.1$} & 
\multicolumn{2}{c|}{$\lambda = 1$} & 
\multicolumn{2}{c}{$\lambda = 10$} \\ \cline{2-9}
& weights & Attributes & weights & Attributes & weights & Attributes & weights & Attributes \\ \hline
\#1& 2.0591 & runq-sz & 1.907 & runq-sz & 1.5179 & runq-sz & 0.1649 & \%sys all \\
\#2& 1.3902 & plist-sz & 1.5266 & plist-sz & 1.3218 & ldavg-1 & 0.1649 & rd\_sec/s dev253-1\\ \#3& 1.2291 & ldavg-1 & 1.3514 & ldavg-1 & 1.2645 & plist-sz & 0.1648 & rxpck/s eth0 \\
\#4& 1.1666 & tcpsck & 1.2505 & tcpsck & 0.9829 & tcpsck & 0.1648 & \%sys 5 \\
\#5& 0.9077 & cswch/s & 0.8944 & cswch/s & 0.94 & cswch/s & 0.1648 & \%sys 2 \\
\#6& 0.8969 & bufpg/s & 0.8323 & bufpg/s & 0.88 & bufpg/s & 0.1647 & totsck \\
\#7& 0.6911 & rxdrop/s eth0 & 0.7129 & totsck & 0.6897 & kbbuffers & 0.1647 & majflt/s \\
\#8& 0.5885 & await dev8-0 & 0.6536 & rxdrop/s eth0 & 0.6383 & totsck & 0.1647 & intr/s \\
\#9& 0.5835 & kbbuffers & 0.6048 & await dev8-0 & 0.5802 & ldavg-15 & 0.1647 & \%iowait all \\
\#10& 0.5635 & kbcached & 0.5586 & proc/s & 0.5789 & await dev8-0 & 0.1646 & tps dev253-1 \\ \hline
\end{tabular}
\end{table*}
The detail results are shown in \ref{tab:selection_L1LS} and \ref{tab:selection_MCL21LS}. We find that the selected attributes are consistent with our expection, indicating that our method are effective and that the selected features are available for further research analysis. For instance, we can analysis the relationship between anomalies and the selected attributes, to prevent system failure and improve cloud computing stability. We also note that MCL21LS selects almost the same attributes as L1LS when $\lambda \in \{0.01, 0.1, 1\}$. This observation gives us an idea that the selected features would not change for different detection tasks and should be consistent. As a further study, We compare the results of our selections with respect to $\lambda$ value. Experimental results show that when $\lambda=10$, the weights of attributes are very close, showing that it is challenged by the fundamental difficulties of attribute selection. On the contrary, when $\lambda \in \{0.01, 0.1, 1\}$, we didn't suffer from the problem. Empirically, we set $\lambda$ to $0.01$ in our experiments.

\section{Conclusion}
In this paper, we present an online self-evolving framework for cloud computing environment anomaly detection.  It introduces self-evolving paradigm to update detector and reduce administrator labeling work. We show that only labeling about 21.06\% data, our self-evolving framework can achieve comparable performance to state-of-art related work. For better performance, we present L1LS detector for type-agnostic anomaly detection and MCL1LS, MCL21LS detectors for type-specific anomaly detection. In our experiments, self-evolving framework tends to yield consistent improvement in both sensitivity and specificity with growing number of could computing records. Moreover, different from previous work, our framework has the capability of selecting the most distinguishing attributes for further study. We expect that a careful re-implementation of our framework and carefully choosing hyper-parameters will further boost the performance.

Combining our self-evolving rule and proposed L1LS/MCL21LS detector, our framework naturally integrates the properties of online learning, labeling reduction, feature selection and anomaly detection. However, our proposed detectors compromise to linear space due to the using of SGD. We plan to explore non-linear kernel projection with our self-evolving framework in future work.

\bibliographystyle{IEEEtran}
\bibliography{Bibliography}

\end{document}